\documentclass{nature}

\usepackage{graphicx}
\usepackage{color}
\usepackage{amsbsy}
\usepackage{mathrsfs}
\usepackage{mathpazo,bm}
\usepackage{amsmath}

\newcommand{\taut}{\tau_{_T}}
\newcommand{\Fig}[1]{Fig.~\ref{#1}}
\newcommand{\fig}[1]{\Fig{#1}}

\newcommand{\apj}{Astrophys.~J.}

\newcommand{\aap}{Astron.~Astrophys.}

\newcommand{\nat}{Nature}

\newcommand{\anrevfm}{Ann.~Rev.~Fluid~Mech.}
\newcommand{\set}[4]{{\it {#1}}, {\bf {#2}}, {#3}, (#4).}
\newcommand{\ttimes}[1]{10^{#1}}
\renewcommand{\v}[1]{{\boldsymbol{#1}}}

\newlength{\singlecolumn}
\newlength{\middlecolumn}
\newlength{\doublecolumn}
\title{Formation of sharp eccentric rings in debris disks with gas but
  without planets}

\author{W. Lyra$^{1,2,3,4}$ \& M. Kuchner$^5$}

\begin{document}

\maketitle

\begin{affiliations}
 \item Jet Propulsion Laboratory, California Institute of Technology, 4800 Oak Grove Drive, Pasadena, CA 91109, USA
 \item Division of Geological \& Planetary Sciences, California Institute of Technology, 1200 E California Blvd MC 150-21, Pasadena, CA 91125 USA
 \item Department of Astrophysics, American Museum of Natural History, 79th Street at Central Park West, New York, NY 10024, USA
 \item Sagan fellow 
 \item NASA Goddard Space Flight Center, Exoplanets and Stellar Astrophysics Laboratory, Code 667, Greenbelt, MD 21230, USA
\end{affiliations}

\begin{abstract}

\noindent{\bf 
``Debris disks'' around young stars (analogues of the
Kuiper Belt in our Solar System) show a variety of non-trivial structures
attributed to planetary perturbations and used to constrain the
properties of the planets\cite{Kuchner2003, chiang2009,
  lagrange2010}. However, these analyses
have largely ignored the fact that some debris disks are
found to contain small quantities of gas\cite{Zuckerman1995,Lagrange1998,Roberge2006,Redfield2007,Maness2008,Moor2011},
a component that all such disks
should contain at some level\cite{Gregorieva2007, Czechowski-Mann2007}.
Several debris disks have been
measured with a dust-to-gas ratio around unity\cite{Zuckerman1995,Lagrange1998,Roberge2006,Redfield2007,Maness2008,Moor2011}
at which the effect of hydrodynamics on the structure of the disk cannot be
ignored\cite{Klahr-Lin2005, Besla-Wu2007}.  Here we report linear and
nonlinear modelling that shows that dust-gas interactions can 
produce some of the key patterns attributed to planets. We find 
  a robust clumping instability that organizes the 
  dust into narrow, eccentric rings, similar to the Fomalhaut debris
  disk\cite{Kalas}. The conclusion that such disks might contain
  planets is not necessarily required to explain these systems.
}

\end{abstract}

Disks around young stars seem to pass through an evolutionary phase when 
the disk is optically-thin and the dust-to-gas ratio $\varepsilon$ ranges from 0.1 to 10.  
The nearby stars $\beta$ Pictoris\cite{Lagrange1998, Olofsson2001, Brandeker2004, Roberge2006, Troutman2011}, HD32297\cite{Redfield2007}, 49 Ceti\cite{Zuckerman1995}, and HD 21997\cite{Moor2011}, 
all host dust disks resembling ordinary debris disks and also have stable 
circumstellar gas detected in molecular CO, Na I or other metal lines; the 
inferred mass of gas ranges from Lunar masses to a few Earth masses
(Supplementary Information, Sect\,1). The gas in 
these disks is thought to be produced by planetesimals or dust grains themselves, 
via sublimation, photodesorption\cite{Gregorieva2007} or collisions\cite{Czechowski-Mann2007}, processes that should occur in every debris 
disk at some level.

Structures may form in these disk via a recently 
proposed instability\cite{Klahr-Lin2005, Besla-Wu2007}. Gas drag causes dust in a 
disk to concentrate at pressure maxima\cite{Take-Arty2001}; 
however, when the disk is optically-thin to starlight, the gas is most likely 
primarily heated by the dust, by photoelectric heating.  In this circumstance, 
a concentration of dust that heats the gas creates a local pressure maximum 
that in turn can cause the dust to concentrate more.   
The result of this {\it photoelectric instability} could be that the dust clumps into rings 
or spiral patterns or other structures, that could be detected via coronographic 
imaging or other methods. 

Indeed, images of debris disks and transitional disks 
show a range of asymmetries and other structures that call for explanation. 
Traditionally, explanations for these structures rely on planetary perturbers -- 
a tantalizing possibility.  But so far, it has been difficult to prove
that these patterns are clearly associated with exoplanets\cite{Janson,Currie}.

Previous investigations of hydrodynamical instabilities in debris disks
neglected a crucial aspect of the dynamics: 
the momentum equations for the dust and gas. Equilibrium
terminal velocities are assumed between time-steps in the numerical 
solution, and the dust distribution is updated accordingly. The
continuity equation for the gas is not solved, i.e., the gas
distribution is assumed to be time-independent, despite heating,
cooling, and drag forces. Moreover, prior investigations only 
considered one-dimensional models, which can only investigate
azimuthally symmetrical ring-like patterns. This limitation also left open the possibility that, in higher dimensions, 
the power in the instability might collect in higher azimuthal
wavenumbers, generating only unobservable clumps.

 We present simulations of the fully compressible problem,
  solving for the continuity, Navier-Stokes, and energy equations for
  the gas, and the momentum equation for the dust. Gas and dust interact dynamically through a
  drag force, and thermally via photoelectric heating. These are
  parametrized via a dynamical coupling time $\tau_f$, and a thermal coupling
  time $\taut$ (Supplementary
  Information, Sect 2). The simulations are performed
  with the Pencil Code\cite{Brandenburg&Dobler,Lyra08,Lyra09, Youdin&Johansen}, which solves the
  hydrodynamics on a grid. Two numerical models are presented: (1) a three-dimensional
  box embedded in the disk that co-rotates with the flow at a fixed
  distance from the star; and (2) a two-dimensional global model of
  the disk in the inertial frame. In the former the dust is treated as
  a fluid, with a separate continuity equation. In the
  latter the dust is represented by discrete particles with position
  and velocities that are independent of the grid.

\begin{figure}
\caption{{\bf Linear analysis of the axisymmetric modes of the photoelectric
  instability.} Solutions for axisymmetric perturbations  
  $\psi^\prime = \hat{\psi}ˆ\exp(st + ikx)$, where $\hat{\psi}$ is a small 
    amplitude, $x$ is the radial coordinate in the local Cartesian 
    co-rotating frame, $k$ is the radial wavenumber, $t$ is
    time and $s$ is the complex frequency. Positive real $s$ means that a
  perturbation grows, negative $s$ indicates that a
  perturbation is damped, and imaginary $s$
  represents oscillations. Solutions are for $\alpha=0$, $\tau_f=1/\varOmega_K$, and $\taut$=0.
  {\bf a-e}, The five solutions as functions of $n=kH$ and $\varepsilon$. Solutions {\bf a-c}
  show linear growth. Growth is restricted to the region with low dust-to-gas ratio ($\varepsilon < 1$),
  high wavenumber ($n > 1$). The growing modes
  in {\bf b} and {\bf c} have non-zero oscillation frequencies, characterizing an overstability. Conversely, solution {\bf a} 
  is a true instability. {\bf d, e}, Solutions that correspond to damped oscillations through most of the
  parameter space. In a small region (high dust-to-gas ratio and high
  frequency), modes are exponentially damped without oscillating. 
  {\bf f, g}, Growth rate ({\bf f}) and oscillation frequency ({\bf
    g}). Using $\xi = \varepsilon n^2$ and taking the limit $\xi \gg 1$  
  permits better visualization of the three behaviours: true instability (red), overstability (yellow), 
  and damped oscillations (blue). The other two solutions are
  the complex conjugate of the oscillating solutions, and are not shown. 
  {\bf h}, Growth rates. When viscosity is considered ($\alpha=\ttimes{-2}$ in this
  example), power is capped at high wavenumber, leading to a finite 
  most-unstable wavelength. The figure shows the analytical prediction of the
  linear instability growth in this case (Supplementary Information,
  Sect 4) compared to the growth rates measured numerically. 
  The overall agreement is excellent. The growth rates are only very
  slightly underestimated. {\bf i}, Nonlinear growth. Although there is no linear
  instability for $\varepsilon=1$, growth occurs when the amplitude of
  the initial perturbation ($u_{\rm rms}$) is increased, a hallmark of nonlinear instability.
\label{fig:linear}
  \begin{center}
      \includegraphics[width=\doublecolumn]{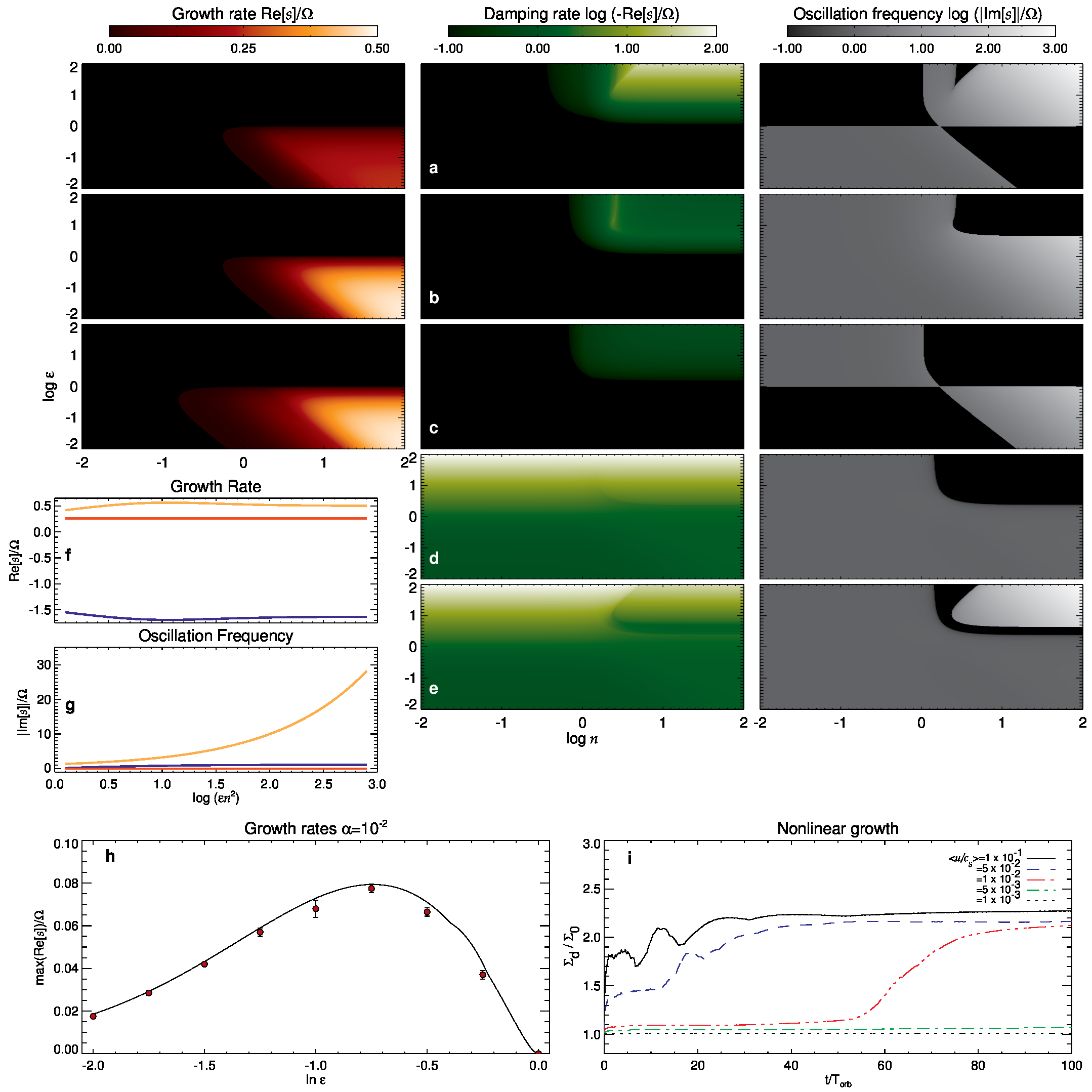}
  \end{center}
}
\end{figure}

  We perform a stability analysis of the linearized system of
  equations, that should help interpret the results of the
  simulations (Supplementary Information, Sect\,3). We plot in \fig{fig:linear}a-c the three solutions that show 
  linear growth, as functions of $\varepsilon$ and $n=kH$, where $k$ is the radial wavenumber and $H$
  is the gas scale height
  ($H$=$c_s/\sqrt{\gamma}\varOmega_K$, where $c_s$ is the sound speed,
  $\varOmega_K$ the Keplerian rotation frequency and $\gamma$ the
  adiabatic index). The friction time $\tau_f$ is assumed to be equal to $1/\varOmega_K$.
  The left and middle panels show the growth and damping rates. The right panels show the 
  oscillation frequencies. There is no linear instability for $\varepsilon \ge 1$ or $n\le 1$. At low dust load and 
  high wavenumber the three growing modes appear. The growing modes 
  shown in \fig{fig:linear}a have zero oscillation frequency, 
  characterizing a true instability. The two other growing solutions 
  (\fig{fig:linear}b-c) are overstabilities, given the 
  associated non-zero oscillation frequencies. The pattern of 
  larger growth rates at large $n$ and low $\varepsilon$ 
  invites to take $\xi=\varepsilon n^2$ as characteristic variable, and explore
  the behavior of $\xi \gg 1$. The solutions in this approximation are 
  plotted in \fig{fig:linear}f-g. The instability (red) 
  has growth rate $\approx$ 0.26\,$\varOmega_K$ for all $\xi$. The overstability
  (yellow) reaches an asymptotic growth rate of $\varOmega_K/2$,
  at ever growing oscillation frequencies. Damped oscillations (blue) occur 
  at frequency close to the epicyclic frequency.

  Whereas the inviscid solution has growth even for very small
  wavelengths, viscosity will cap power at this regime, leading to a
  finite fastest growing mode (Supplementary Information,
  Sect\,4), 
which we reproduce numerically (\fig{fig:linear}h). Although there is no
  linear growth for $\varepsilon \ge 1$, we show that there
  exists nonlinear growth for $\varepsilon=1$. We show in
  \fig{fig:linear}i the time evolution of the maximum dust surface
  density $\varSigma_d$ (normalized by its initial value,
  $\varSigma_0$). A qualitative change in the behaviour of the system
  (a bifurcation) occurs when the noise amplitude of the initial
  velocity ($u_{\rm rms}$) is raised far enough, as expected from nonlinear
  instabilities\cite{Stuart,Lesur&Papaloizou}. We emphasize this
  result because, depending on the abundance of H$_{\rm 2}$, the
  range of $\varepsilon$ in debris disks spans both the
  linear and nonlinear regimes.  The parameter space of $\taut$ and $\tau_f$ 
  is explored in one-dimensional models in the Supplementary Information (Sect\,5), showing robustness.

\begin{figure}
\caption{{\bf Growth and saturation of the photoeletric instability.} In this threedimensional stratified
local box with linearized Keplerian shear, the main source of heating
is photoelectric. The equilibrium in the radial direction is between
stellar gravity, Coriolis force, and centrifugal force. In the vertical direction
the equilibrium for the gas is hydrostatic, between stellar gravity,
pressure, and the drag-force back-reaction. To provide a stable stratification,
an extra pressure $p_b=\rho c_b^2$ is added, where $c_b$ is a sound speed associated with 
a background temperature. For the dust, a steady state is established
between gravity, diffusion, and drag force. The dust
continually falls to the midplane but is diffused
upwards. The diffusion is applied only in $z$, mimicking turbulent diffusion that is in general
anisotropic. {\bf a}, $x-z$ cut at $y=0$ at 100 orbits. 
The instability concentrates dust in a
preferred wavelength. The resulting structures have stable
stratification. {\bf b}, $x-y$ cut at the midplane $z=0$ at 100
orbits. No non-axisymmetric instability is observed, and the dust 
forms stripes. {\bf c}, Time-evolution of the vertically and azimuthally averaged
density, showing the formation of well-defined rings. {\bf d}, Time
evolution of the maximum dust density. The instability saturates at 
$\approx$70 orbits in this case. The slowdown compared with 
the growth rate $\varOmega_K/2$ predicted in \fig{fig:linear} is because of the
use of viscosity, and the background pressure needed 
for the stratification. The dimensionless parameter $\beta =
\gamma(c_b/c_s)^2$ measures the strength of this term. {\bf e} Maximum
growth rate, showing that linear
instability exists as long as $\beta < 1$. The maximum growth rates drecreases smoothly from $\varOmega_K/2$
for $\beta=0$, to zero for $\beta=1$. {\bf f}, The structure
formed in the dust density at $t=50$ (about 8 orbits) 
for different values of $\beta$. At moderate
values, growth still occurs at a significant fraction of the
dynamical time. The run shown in panels {\bf a-d} used $\beta$=0.5.
\label{fig:strat}
  \begin{center}
      \includegraphics[width=\doublecolumn]{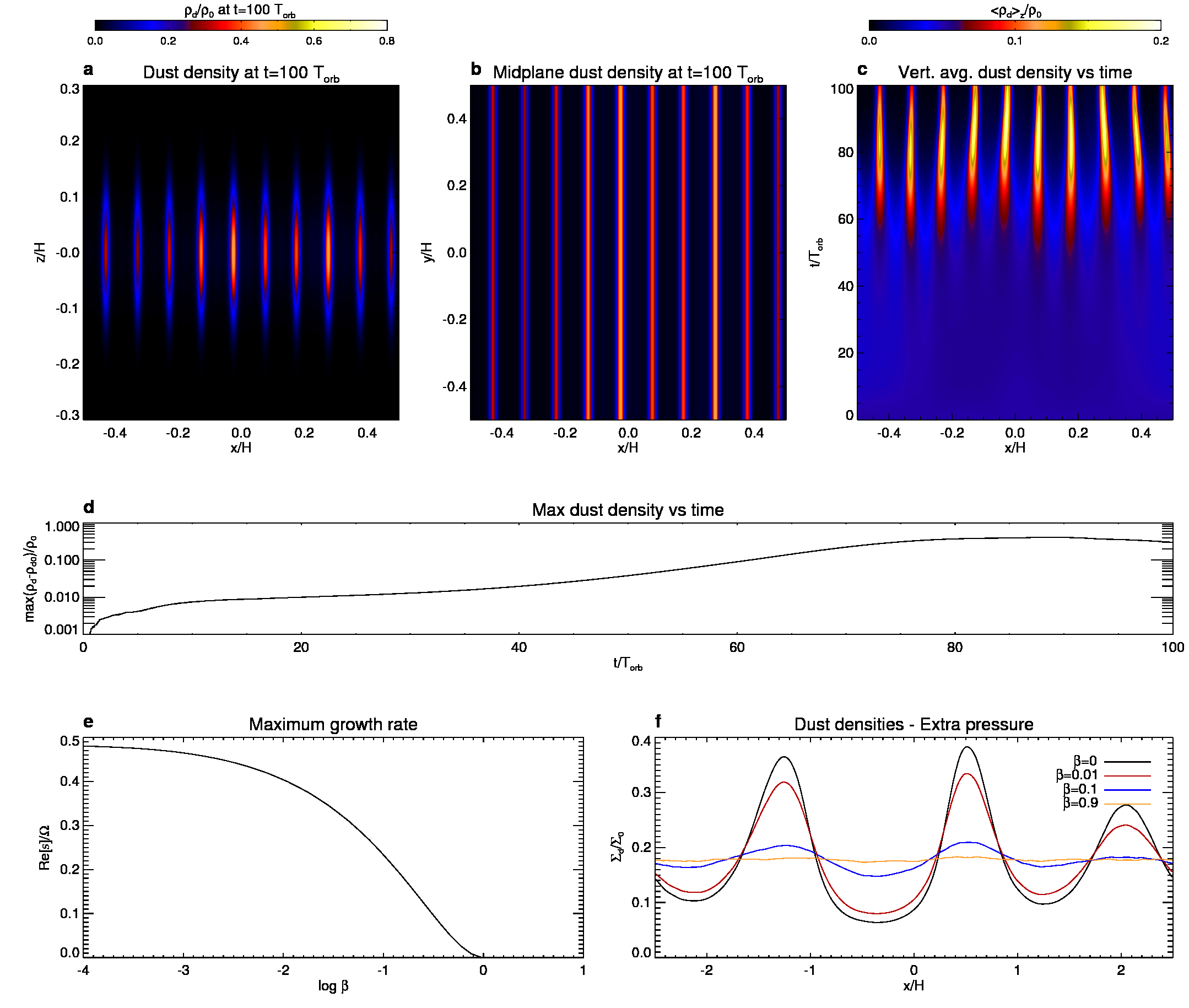}
  \end{center}
}
\end{figure}

  In \fig{fig:strat} we show the linear development and saturation of the 
  photoelectric instability in a vertically stratified local box of size $(1 \times 1 \times 0.6)H$ 
  and resolution 255 $\times$ 256 $\times$ 128. The dust and gas are
  initialized in equilibrium (Supplementary Information,
  sect\,6). The dust-to-gas ratio is given by ${\rm log}\,\varepsilon=-0.75$, so
  that there is linear instability, and viscosity $\nu = \alpha c_s H$ is applied as 
  $\alpha=\ttimes{-4}$ (where $\alpha$ is a dimensionless parameter\cite{Shakura&Sunyaev}). 
  The initial noise is $u_{\rm rms}/c_s = \ttimes{-2}$. \fig{fig:strat}a
  shows the dust density in the $x$-$z$ plane, and \fig{fig:strat}b in the $x$-$y$ plane, 
  both at 100 orbits (the orbital period is $T_{\rm
    orb}=2\pi/varOmega_K$). 
  \fig{fig:strat}c shows the one-dimensional $x$-dependent vertical and azimuthal average against time.
   Through photoelectric heating, pressure maxima are generated at the locations where dust
  concentrates, that in turn attract
  more dust by means of the drag force. There is no hint of unstable 
  short-wavelength (less than $H$) non-axisymmetric modes: the instability generates
  stripes. The simulation also shows that
  stratification does not quench the instability. \fig{fig:strat}d
  shows a plot of the maximum dust density against time, achieving saturation and steady
  state at about 70 orbits.

\begin{figure}
\caption{
{\bf Sharp eccentric rings.}  
{\bf a-c}, Snapshots of the dust density in a two-dimensional global
disk in polar coordinates, at 20 orbits ({\bf a}), 40 orbits ({\bf b}),
and 60 orbits ({\bf c}). The photoelectric instability
initially concentrates the dust axisymmetrically into rings, at 
a preferred wavelength. As the simulation proceeds, some rings maintain the axisymmetry, others break into arcs. Some arcs
rearrange into rings at later times, such as those at $r=0.6$ and
$r=1.0$ between {\bf b} and {\bf c}. Alhough mostly
axisymmetric, some rings seem to oscillate, appearing off-centered
or eccentric. {\bf d}, We measure the azimuthal spectral power of the density
shown in {\bf c}, as a function of radius. Modes from $m$=0 to $m$=3
are shown, where $m$ is the azimuthal wavenumber. {\bf e}, Although the ring at
$r=1.5$ has $m=0$ as the more prominent mode, we show that a 
circle (black dotted line) is not a good fit. An ellipse
of eccentricity $e=0.03$ (red dotted line) is a better fit, although still falling short of
accurately describing its shape. The black and red diamonds are the
centre of the circle (the star), and the centre of the ellipse (a
focal distance away from the star), respectively. 
\label{fig:rings}
  \begin{center}
      \includegraphics[width=\doublecolumn]{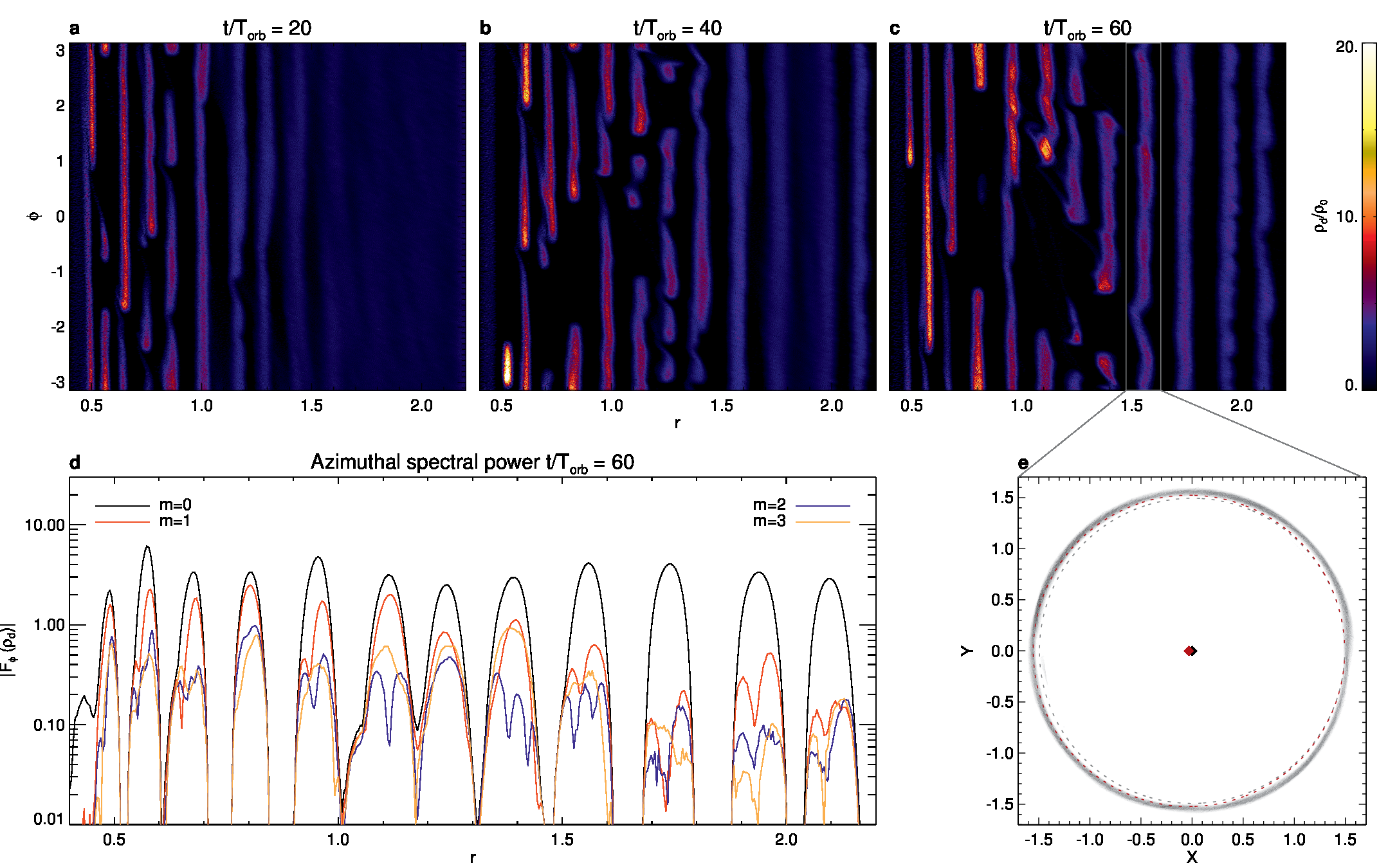}
  \end{center}
}
\end{figure}

  We consider now a two-dimensional global model. The resulting flow, in the
  $r-\phi$ plane ($r$ is radius and $\phi$ is azimuth), is shown in \fig{fig:rings}a-c at selected
  snapshots. The flow develops into a dynamic system of narrow
  rings. Whereas some of the rings break into arcs, some maintain
  axisymmetry for the whole timespan of the simulation. It is also
  observed that some arcs later re-form into rings. We check that in the absense of the
  drag force back-reaction, the system does not develop rings 
  (Supplementary Information, Sect 7). We also check that when the
  conditions for the streaming instability\cite{Youdin&Johansen} are considered, the
  photoelectric instability dominates (Supplementary Information,
  Sect 8).

A development of the model is that some of the rings start to
oscillate, seeming eccentric. These oscillations are epicycles in the orbital plane, with a period
equaling the Keplerian, 
corresponding to the free oscillations in the right hand side of
\fig{fig:linear}a-c. We check 
(Supplementary Information, Sect 9) that they correspond to eigenvectors for which $\v{u}$ = $\v{v}$, that is, gas and
dust velocities coinciding. For this mode, the drag force and
back-reaction are cancelled. So, for maintaining the eccentricity, this
mode is being selected among the other modes in the spectrum. This is
naturally expected when the dust-to-gas ratio is very high. For
$\varepsilon \gg 1$, the gas is strongly coupled to the dust,
canceling the gas-dust drift velocity in the same way that $\tau_f \ll
1$ does in the opposite way, by strongly coupling the dust to the
gas. In this configuration, the freely oscillating epicyclic modes can be selected. 

We plot in \fig{fig:rings}e one of the oscillating rings,
showing that its shape is better fit by an ellipse (red dotted line)
than by a circle (black dotted line). The eccentricity is 0.03, which is close to the 
eccentricity found\cite{Buenzli} for the ring around 
HD 61005 ($e$=$0.045 \pm 0.015$). We also notice 
 that some of the  clumps in \fig{fig:rings} should become very bright in reflected light, as they have dust
  enhancements of an order of magnitude. In conclusion, the proposed 
photoelectric instability provides simple and plausible explanations
for rings in debris disks, their eccentricities, 
and bright moving sources in reflected light.

Recent work\cite{Boley} suggests that the ring around Fomalhaut is confined by a pair 
of shepherding terrestrial-mass planets, below the current detection 
imits. Detection of gas around the ring would be a way to distinguish 
that scenario from the one we propose. At present, only upper limits on 
the amount of gas in the Fomalhaut system exist\cite{Liseau1999};
however, they are relatively insensitive because they probe CO emission, and CO
could easily be dissociated around this early A-type star. 

\newpage

\newpage

\newpage

\begin{addendum}
 \item [Supplementary Information] is linked to the online version of the 
   the paper at www.nature.com/nature
 \item [Acknowledgments] We thank H. Latter and G. Stewart for discussions. The writing of this paper started at the American Museum of Natural History, with financial support by the National Science Foundation under grant no. AST10-09802, and was completed at the Jet Propulsion Laboratory, California Institute of Technology, under a contract with the National Aeronautics and Space Administration. This research was supported by an allocation of advanced computing resources supported by the National Science Foundation. The computations were performed on the Kraken system at the National Institute for Computational Sciences. W.L. is a Carl Sagan fellow. M.K. is supported in part by the NASA Astrobiology Institute through the Goddard Center for Astrobiology.
 \item [Author Contributions] W.L. contributed to developing the model, 
performed the calculations, and wrote the manuscript. M.K.  contributed 
to developing the model and writing the manuscript.
\item [Author information] Reprints and permissions information are available 
at www.nature.com/reprints. The authors declare no competing financial interests. 
Readers are welcome to comment on the online version of the
paper. Correspondence and requests for materials should be addressed to W.L. (wlyra@caltech.edu) or
M.K. (marc.j.kuchner@nasa.gov).
\end{addendum}

\end{document}